\documentclass[showpacs, oneside, twocolumn, prl, amsmath, amssymb, nofootinbib, superscriptaddress]{revtex4-1}

\usepackage{amsmath,amssymb,amsfonts,array}
\usepackage{bm,bbm}
\usepackage{cases}
\usepackage{dcolumn}
\usepackage{graphicx}
\usepackage{hyperref}
\usepackage{subfigure}
\usepackage{wasysym}
\usepackage{xcolor}
\usepackage{cancel}

\definecolor{dullred}{rgb}{0.706,0.208,0.192}
\definecolor{darkred}{rgb}{0.545,0,0}
\definecolor{MaroonC}{rgb}{0,0.502,0.502}
\definecolor{dullblue}{rgb}{0,0.298,0.49}
\definecolor{blue3}{RGB}{31, 119, 180}
\definecolor{dullpurple}{rgb}{0.431,0.188,0.534}
\definecolor{darkgreen}{rgb}{0.075,0.302,0.047}
\definecolor{darkergreen}{rgb}{0,0.196,0.125}
\definecolor{darkergreen2}{rgb}{0,0.294,0.188}

\hypersetup{colorlinks=true,
	breaklinks=true,
	pdfstartview=Fit,
	linkcolor=blue,
	citecolor=blue,
	urlcolor=blue}

\def\be{\begin{equation}}
\def\ee{\end{equation}}
\def\ba{\begin{eqnarray}}
\def\ea{\end{eqnarray}}

\def\nn{\nonumber}
\def\lf{\left}
\def\rt{\right}

\def\d{{\mathrm{d}}}

\begin{document}
	

\title{Null energy condition violation and beyond Horndeski physics in light of DESI DR2 data}

\author{Gen Ye}
\email[]{ye@lorentz.leidenuniv.nl}
\affiliation{Institute for Astrophysics, School of Physics, Zhengzhou University, Zhengzhou 450001, China}
\affiliation{Institute Lorentz, Leiden University, PO Box 9506, Leiden 2300 RA, The Netherlands}
\author{Yong Cai}
\email[Corresponding author:~]{caiyong@zzu.edu.cn}
\email[]{yongcai\_phy@outlook.com}
\affiliation{Institute for Astrophysics, School of Physics, Zhengzhou University, Zhengzhou 450001, China}

\begin{abstract}

Inflation and dark energy (DE), both featuring accelerated expansion, are crucial components of modern cosmology. As indicated by singularity theorems, null energy condition (NEC) violation is essential for resolving the initial singularity of inflation. The latest DESI DR2 results show that the DE equation of state evolves from $w_{\rm DE} < -1$ at redshift $z \lesssim 1$ to $w_{\rm DE} > -1$ today, reinforcing interest in testing NEC violation in the observable Universe. Within a unified effective field theory framework that applies to both inflation and DE, we use DESI DR2 data and a non-parametric reconstruction approach to test, for the first time, the ``beyond Horndeski'' physics known to support fully stable NEC violation in nonsingular cosmology. Therefore, our work provides a novel avenue to constrain the physics underlying NEC violation in a model agnostic way. It also highlights the potential relevance of NEC violation to both the primordial Universe and late-time cosmic acceleration.

\end{abstract}

\maketitle



{\it Introduction.---} Modern cosmology faces numerous mysteries. At one end of time, precise observations of the initial state of our Universe through the cosmic microwave background (CMB) \cite{Planck:2019nip,ACT:2025fju} provide evidence for a period of accelerated expansion in the very early Universe, known as inflation \cite{Guth:1980zm,Starobinsky:1980te,Linde:1981mu,Albrecht:1982wi}. At the other end of time, observations of type Ia supernovae (SNIa) show that the present Universe is also undergoing accelerated expansion, which is believed to be driven by dark energy (DE). Furthermore, the CMB observes an initial perturbation spectrum that is not exactly scale-invariant \cite{Planck:2019nip,ACT:2025fju}, indicating that the physics driving inflation should be dynamical. Given this, it is perhaps not surprising that the DE driving the current accelerated expansion has also recently been observed to be dynamical \cite{DESI:2024lzq,DESI:2024uvr,DESI:2024mwx,DESI:2025zgx,DESI:2025zpo}. However, the physical nature of the inflaton and DE, and whether they share the same underlying physics \cite{Brandenberger:2025hof}, remains a mystery.


Inflation inevitably encounters a cosmological singularity in the finite past \cite{Borde:2001nh}, presenting serious theoretical challenges for explaining its initial conditions. The null energy condition (NEC), which requires the energy-momentum tensor $T_{\mu\nu}$ to satisfy $T_{\mu\nu}n^{\mu}n^{\nu} \geq 0$ for any null vector $n^\mu$, is one of the key assumptions underlying the singularity theorem \cite{Penrose:1964wq}.
Cosmologies that evade the singularity problem are referred to as nonsingular cosmologies, typically featuring a period of NEC violation (see \cite{Rubakov:2014jja} for a review) either prior to the onset of inflation or at other stages.  Intriguingly, recent observation of the baryon acoustic oscillation (BAO) by the DESI group indicates that the DE in the observed Universe might also have experienced an NEC violating period near the onset of DE domination, i.e., the DE equation of state parameter $w_{\rm DE}$ is observed to cross the phantom divide from $w_{\rm DE}<-1$ ($T^{(\rm DE)}_{\mu\nu}n^\mu n^\nu=\rho_{\rm{DE}} + P_{\rm{DE}}=(1+w_{\rm{DE}})\rho_{\rm{DE}}<0$) to the present $w_{\rm DE}>-1$ at $z<1$ \cite{DESI:2025fii}, see also \cite{Feng:2004ad,Wei:2005nw,Caldwell:2005ai,Cai:2007zv}. This discovery not only elaborates the similarity between the opposite ends of the cosmic evolution, but also opens up \textit{the first window} to study NEC violating physics using cosmological observations.

The primordial Universe and the present-day Universe are not only qualitatively similar to some extent but also theoretically unified within the same effective field theory (EFT) framework. The EFT approach provides a general way of studying new physics in any physical system with a given set of symmetries by building the most general description of the system consisting of all possible operators compatible with the assumed symmetry. The fundamental nature of cosmology is the time-dependent cosmological evolution, which points to the EFT assuming time-dependent spatial diffeomorphism symmetry with broken time diffeomorphism invariance. The resulting EFT action up to quadratic order in perturbations, neglecting higher-order spatial derivatives not relevant on cosmological scales, is
	\ba
	S&=&\int
	\d^4x\sqrt{-g}\Big[ {M_p^2\over2} f(t)R-\Lambda(t)-c(t)g^{00}
	\nn\\
	&\,&\qquad\qquad\quad +{M_2^4(t)\over2}(\delta g^{00})^2-{m_3^3(t)\over2}\delta
	K\delta g^{00}
    \nn\\
	&\,&\qquad\qquad\quad
    -m_4^2(t)\lf( \delta K^2-\delta K_{\mu\nu}\delta
	K^{\mu\nu} \rt)
	\nn\\
	&\,&\qquad\qquad\quad  + {\tilde{m}_4^2(t)\over
		2}\delta g^{00}R^{(3)}\Big] +S_{\rm m}[g_{\mu\nu},\psi_{\rm m}]\,,
	\label{action01}
	\ea
see the Supplemental
Material for definition of the included operators. The background and linear perturbation of Horndeski gravity \cite{Horndeski:1974wa}, which covers a wide range of modified gravity theories, is fully described by action \eqref{action01} with $m_4^2=\tilde{m}_4^2$. Physics beyond Horndeski refers to $m_4^2\ne\tilde{m}_4^2$. The EFT \eqref{action01} unifies the description of the NEC violation and inflation in the primordial Universe \cite{Creminelli:2006xe,Cheung:2007st,Weinberg:2008hq} as well as DE \cite{Gubitosi:2012hu,Bloomfield:2012ff,Gleyzes:2013ooa} in the current observable Universe.

The NEC is satisfied by all known forms of matter at the classical level. The physical feasibility of NEC violation has long been a challenge \cite{Carroll:2003st,Cline:2003gs,Dubovsky:2005xd,Nicolis:2009qm,Creminelli:2010ba,Creminelli:2012my,Moghtaderi:2025cns}, as it is often accompanied by catastrophic instabilities of the perturbations, including the ghost and gradient instabilities \cite{Carroll:2003st,Cline:2003gs,Easson:2011zy,Battarra:2014tga,Koehn:2015vvy,Qiu:2015nha,Ijjas:2016tpn,Ijjas:2016vtq,Dobre:2017pnt}. Furthermore, in the study of nonsingular cosmology, it is proved by the ``No-Go" theorem \cite{Libanov:2016kfc, Kobayashi:2016xpl} that a {\it fully stable NEC violation}\footnote{By ``fully stable NEC violation'', we actually mean that no ghost or gradient instabilities are induced by the NEC violation at any time in the geodesically complete cosmic history--not just during the NEC-violating phase itself. In fact, the cubic Galileon term in Horndeski theory is sufficient to shift instabilities from the NEC-violating phase to other phases \cite{Easson:2011zy,Ijjas:2016tpn}, but it cannot completely eliminate them in nonsingular cosmology.} can hardly be achieved within the Horndeski theory \cite{Horndeski:1974wa,Deffayet:2011gz,Kobayashi:2011nu}, which covers a wide range of modified gravity theories. It is first explicitly demonstrated in \cite{Cai:2016thi}, based on \eqref{action01}, that a fully stable NEC violation can be achieved through ``beyond Horndeski'' physics, characterized by the operator $\delta g^{00}R^{(3)}$.\footnote{The same result was independently obtained in \cite{Creminelli:2016zwa}, which was uploaded to arXiv two days later.} The first covariant ``beyond Horndeski'' scalar-tensor theory used to achieve fully stable NEC violation in this regard was later obtained in \cite{Cai:2017dyi},\footnote{A similar finding was also made in Ref. \cite{Kolevatov:2017voe}, which was submitted to arXiv nine days later, see also \cite{Cai:2017dxl,Cai:2017pga,Mironov:2018oec,Ye:2019frg,Ye:2019sth,Mironov:2019qjt,Akama:2019qeh,Mironov:2019mye,Ilyas:2020qja,Ilyas:2020zcb,Zhu:2021whu,Zhu:2021ggm,Cai:2022ori,Akama:2025ows,Dehghani:2025udv}.} and it belongs to the GLPV class \cite{Gleyzes:2014dya,Gleyzes:2014qga}. Potential observable signatures of NEC violation have been discussed in primordial gravitational waves (GWs) \cite{Cai:2020qpu,Cai:2022nqv,Jiang:2023gfe,Ye:2023tpz,Pan:2024ydt}, enhanced parity-violating effects \cite{Cai:2022lec,Jiang:2024woi}, primordial black hole formation and scalar-induced GWs \cite{Cai:2023uhc}.

In this letter, we use the DESI DR2 results to examine the ``beyond Horndeski'' physics within the framework of the EFT by performing non-parametric reconstruction to the EFT functions. The EFT is theoretically quite general in that it is based on symmetry and broadly encompasses various gravity theories, while the reconstruction method is free from bias induced by parametrization. Our result confirms the theoretical expectation that $\delta g^{00}R^{(3)}$ is able to stabilize the NEC violation indicated by DESI. Moreover, we found that, when combining DESI DR2 data with the CMB and SNIa observations, there is a $2\sigma$ hint for the non-trivial existence of $\delta g^{00}R^{(3)}$. Our work not only provides a novel and model-independent avenue to constrain the physics underlying NEC violation through cosmological observations, but also establishes its viability by applying it to the current data.

{\it Physics underlying fully stable NEC violation.---} To gain theoretical insights into the physics, we first briefly review the stability analysis in a nonsingular primordial Universe. In unitary gauge, $h_{i j}=a^2 e^{2\zeta}\left(e^\gamma\right)_{i j}$, where $\gamma_{i i}=0=\partial_i \gamma_{i j}$, the quadratic action of tensor and scalar perturbations can be given as
\be S^{(2)}_{\gamma}={M_p^2\over8}\int \d^4x\,a^3 Q_{\rm T}\lf[
\dot{\gamma}_{ij}^2 -c_{\rm T}^2{(\partial_k\gamma_{ij})^2\over
	a^2}\rt]
\ee
and
\be
S_{\zeta}^{(2)}=\int \d^4x\,a^3 Q_{\rm s}\lf[\dot{\zeta}^2-c_{\rm s}^2 {(\partial\zeta)^2\over a^2} \rt]\,,
\ee
respectively. The derivation can be found in \cite{Cai:2016thi}. We need to ensure that
\be
Q_{\rm T}>0\,,\quad Q_{\rm s}>0\,,\quad c_{\rm T}^2>0\,,\quad c_{\rm s}^2>0
\ee
throughout the entire evolution history of the Universe to avoid ghost instabilities and gradient instabilities.

From the action (\ref{action01}), we have
\ba &\,& Q_{\rm T}= f+{2 m_4^2\over M_p^2}\,, \quad c_{\rm T}^2=f/Q_{\rm T}\,,\label{eq:QTcT02}\\ &\,& Q_{\rm s}=c_1\,,\quad c_{\rm s}^2=\lf({\dot{c}_3\over a}-c_2 \rt)/c_1 \,,\label{eq:QTcT01}
\ea
where
\ba
&\,&
c_1=\frac{Q_{\rm T}}{4 \gamma ^2 M_p^2}
 \Big[
   2 M_p^4 Q_{\rm T}\dot{f} H
   \nn\\ &\,&\qquad\qquad\qquad
   -2 M_p^2 Q_{\rm T} \left(  2 f M_p^2 \dot{H}+\ddot{f}M_p^2
      -4 M_2^4\right)
   \nn\\&\,&\qquad\qquad\qquad
   -6 \dot{f}M_p^2 m_3^3
   +3 \dot{f}^2 M_p^4
   +3 m_3^6\Big]
    \,, \label{c1}\\
&\,&c_2=f M_p^2\,,\label{c2}
\\&\,&
c_3={aM_p^2\over \gamma} Q_{\rm T} Q_{\tilde{m}_4} \,, \label{c3}
\\ &\,&\gamma = HQ_{\rm T}-{m_3^3\over 2M_p^2}+{1\over2}\dot{f}\,, \quad
Q_{\tilde{m}_4}=f+{2\tilde{m}_4^2\over M_p^2}\,. \label{gamma}
\ea
It is not difficult to see that $Q_{\rm T}>0$ and $c_{\rm T}^2>0$ are easy to guarantee. For example, when $f=1$ and $m_4^2=0$, we have $Q_{\rm T}=1$ and $c_{\rm T}=1$, which are the same as those in general relativity.
Due to the complex form of $c_1$, it is clear that ensuring $Q_{\rm s} = c_1 > 0$ is also straightforward by appropriately choosing $M_2^4$ or $m_3^3$.

Therefore, under the conditions of $Q_{\rm T} > 0$, $c_{\rm T}^2 > 0$, and $Q_{\rm s} > 0$, the real challenge lies in achieving $c_{\rm s}^2 > 0$, i.e., ensuring that
\be
\dot{c}_3 >a c_2 \label{eq:c3c202}
\ee
throughout, where $c_2 = f M_p^2 > 0$ since $Q_{\rm T} > 0$ and $c_{\rm T} > 0$.
One possibility is to consider the case where $\int a c_2 \d t$ converges, in which case potential strong coupling issues need to be handled with care (see e.g., \cite{Ijjas:2016vtq,Ageeva:2018lko,Ageeva:2020gti,Ageeva:2021yik,GilChoi:2025hbs}), but we do not discuss this issue further here.

We focus on the case where $\int a c_2 \d t$ does not converge, such as when $f=1$ and $m_4^2=0$ (i.e., $Q_{\rm T}=1$ and $c_{\rm T}=1$). In fact, it is reasonable to consider this case, as no deviations from general relativity have been observed in GWs so far.
In this case, $\sigma = \int a \, \d t$ is the affine parameter of the null geodesics \cite{Borde:2001nh}.
As a result, the geodesic completeness and Eq. (\ref{eq:c3c202}) imply that $c_3$ must cross zero at some point.\footnote{Note that if the geodesic completeness of the Universe is not required, the temporary NEC violation within Horndeski theory is not necessarily forbidden by the ``No-Go'' theorem \cite{Libanov:2016kfc,Kobayashi:2016xpl}.} From Eq. (\ref{c3}), it is evident that this challenging task can only be accomplished by $Q_{\tilde{m}_4}=f+\frac{2\tilde{m}_4^2}{M_p^2}$. This highlights the crucial role of the ``beyond Horndeski'' physics represented by the operator $\delta g^{00}R^{(3)}$ in addressing the issue of perturbation instability.

Interestingly, the operator $\delta g^{00}R^{(3)}$ neither affects the background evolution nor influences tensor perturbations at the quadratic perturbation level.
The covariant form of the operator $\delta g^{00}R^{(3)}$ is obtained in \cite{Cai:2017dyi} (see Supplemental Material), where it provides the first covariant action that can achieve fully stable NEC violation. Due to the vast difference in energy scale, terms that are excited in the primordial Universe are usually irrelevant in DE. This is, however, not the case for the $\delta g^{00}R^{(3)}$ (and the $m_4^2$ term as well) which has the same dimension as the Einstein-Hilbert term. This means that $\delta g^{00}R^{(3)}$ can be relevant in both the primordial and late-time Universe.
In the scenario for the entire cosmic history we consider below, we assume that the scalar field in the covariant form of $\delta g^{00}R^{(3)}$ retains a non-vanishing value after reheating---following its role in driving the nonsingular primordial Universe with NEC violation---and subsequently acts as DE at late times. Under this assumption, the primordial and the late-time Universe can be connected by the same EFT operators.




{\it Methodology and data.---} As explained previously, both the $(\delta g^{00})^2$ and $\delta g^{00}R^{(3)}$ operators are only relevant on the perturbation level, we therefore parametrize the background DE evolution using the Chevallier-Polarski-Linder (CPL) form \cite{Chevallier:2000qy,Linder:2002et}
\begin{equation} \label{eq:cpl}
     w_{\text{DE}}(a) = w_0 + w_a(1-a).
\end{equation}
which has been shown to be sufficient in consistently capturing the observed DE effect \cite{DESI:2025fii}. To demonstrate the effect of the operator $\delta g^{00}R^{(3)}$ in stabilizing NEC  violation of DE, we focus on $\delta g^{00}R^{(3)}$ and $(\delta g^{00})^2$ in this Letter and set $m^3_3=m^2_4=0$ and $f=1$.
Namely, the simplified EFT action we consider in the following analysis is
\ba
&\,&S=\int
\d^4x\sqrt{-g}\Big[ {M_p^2\over2} R-\Lambda(t)-c(t)g^{00}
\nn\\
&\,&\qquad +{M_2^4(t)\over2}(\delta g^{00})^2  + {\tilde{m}_4^2(t)\over
	2}\delta g^{00}R^{(3)}\Big] +S_{\rm m}.
\label{action0328}
\ea
As has been explained previously, to evade both the ghost and gradient instabilities, $\delta g^{00}R^{(3)}$ need to be combined with either $(\delta g^{00})^2$ or $\delta g^{00}\delta K$. Here, we keep only $(\delta g^{00})^2$ because it does not impact observations \cite{deBoe:2024gpf,Ye:2024ywg,Frusciante:2018jzw,Raveri:2019mxg}. According to Eq. (\ref{eq:QTcT02}), the action (\ref{action0328}) yields $c_{\rm T}=1$.

\begin{figure}[htbp]
\includegraphics[width=0.44\textwidth]{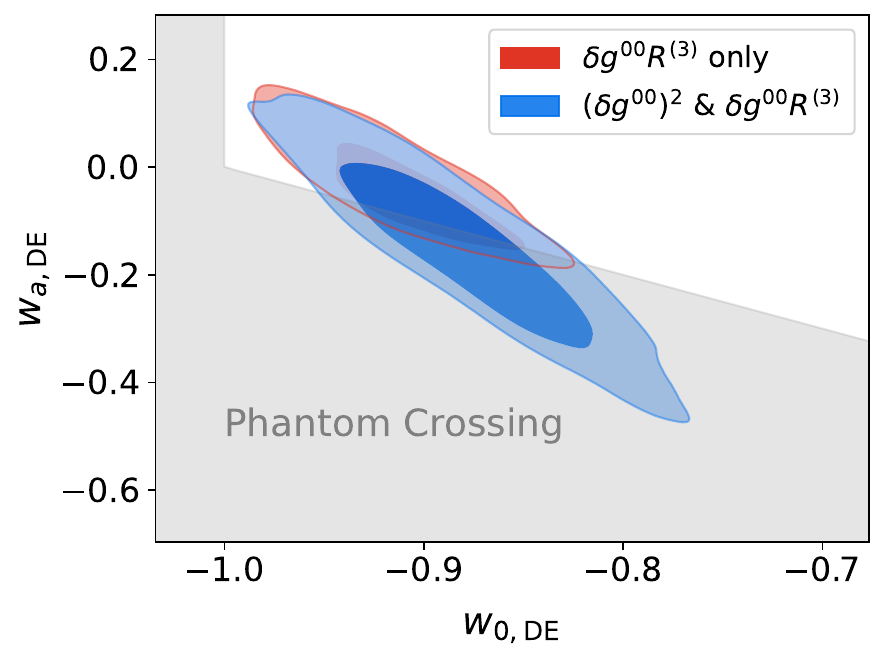}
\includegraphics[width=0.48\textwidth]{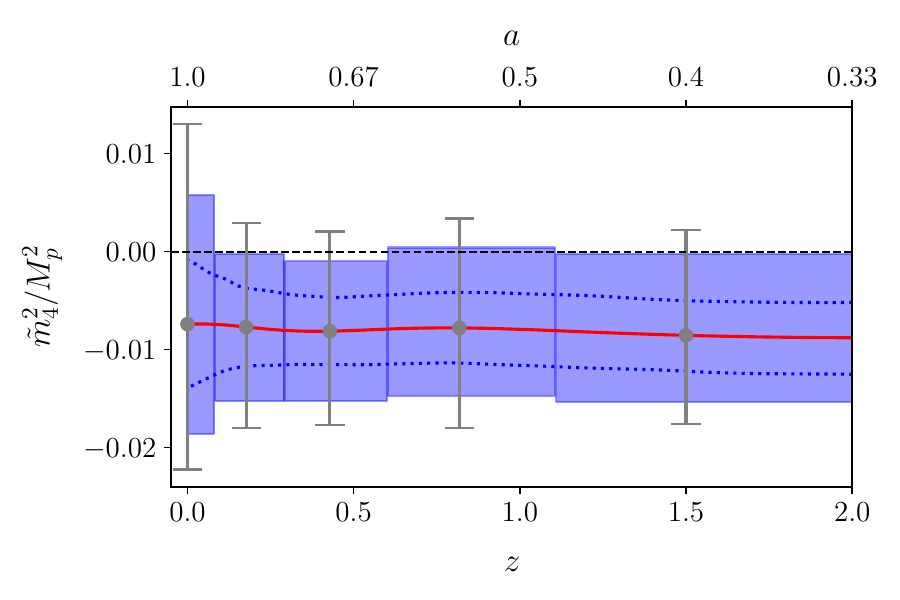}
\caption{\textit{Upper panel:} $1\sigma$ and $2\sigma$ posterior constraints in the $w_0-w_a$ plane of the EFT models with only $\delta g^{00}R^{(3)}$ (red) and both $\delta g^{00}R^{(3)}$ and $(\delta g^{00})^2$ turned on. The gray region indicates the $w_0-w_a$ parameter space where phantom crossing happens. \textit{Lower panel:} the reconstructed functional shape of $\tilde{m}_4^2(a)$, the EFT coefficient of $\delta g^{00}R^{(3)}$. The blue dotted lines mark the $1\sigma$ posterior of $\tilde{m}_4^2(a)$ at each $a$. The gray points are the reconstruction nodes $\{\tilde{m}_{4,i}\}$ with $2\sigma$ range marked by blue boxes and $3\sigma$ range indicated by error bars. Red line plots the mean function of $\tilde{m}_4^2(a)$. The horizontal black dashed line marks the general relativity limit $\tilde{m}_4^2=0$. }
\label{fig01}
\end{figure}

The stability analysis in the previous section assumes the scalar field is the only propagating scalar degree of freedom (DoF). In the real cosmological scenario relevant to DE, matter is present and can propagate additional scalar DoFs. This will modify the stability conditions for DE due to mixing between matter and gravity \cite{Kase:2014yya,Gleyzes:2014dya,DeFelice:2016ucp,Frusciante:2016xoj}. According to Ref. \cite{DeFelice:2016ucp}, when neglecting radiation in the late-time Universe, the sound speed of the DE field described by \eqref{action0328} with matter presence in the sub-horizon limit is
\begin{equation}
\begin{aligned}
c_{\rm s}^2&=\Big[(1+\alpha_{\rm H})(1+w_{\rm DE})(1-\Omega_{\rm m})-\alpha_{\rm H}(1+2\alpha_{\rm H})\Omega_{\rm m}
\\
&\quad +\frac{2}{3}\alpha_{\rm H} +\frac{2}{3}\frac{{\rm d}\alpha_{\rm H}}{{\rm d}\ln a}\Big]\cdot\frac{3M_p^2}{2Q_{\rm s}}\,,
\end{aligned}
\end{equation}
%
where $Q_{\rm s}=\frac{3M_p^2}{2}(1+w_{\rm DE})(1-\Omega_{\rm m})+\frac{2M_2^4}{H^2}$. We have defined $\rho_{\rm DE}\equiv3M_p^2H^2-\rho_{\rm m}$, $\rho_{\rm DE}+P_{\rm DE}\equiv(1+w_{DE})\rho_{\rm DE}\equiv-2M_p^2 \dot{H}-(\rho_{\rm m}+P_{\rm m})$, $\Omega_{\rm m}(t)\equiv\frac{\rho_{\rm m}(t)}{3M_p^2H^2(t)}$ and $\alpha_{\rm H}\equiv2\tilde{m}_4^2/M_p^2$. Eq. \eqref{eq:QTcT01} is recovered in the limit $\Omega_{\rm m}\to0$. Assuming $\alpha_{\rm H}\simeq {\rm const.}$, to avoid gradient instability when $\Omega_{\rm m} \simeq 1$ and $1 + w_{\rm DE} < 0$, it is required that $-1/6 \lesssim \alpha_{\rm H} \lesssim 0$, or equivalently $-1/12 \lesssim \tilde{m}_4^2 / M_p^2 \lesssim 0$, while ghost instability can be avoided by ensuring $M_2^4 > 0$. Having non-zero $\alpha_{\rm H}$ and $M_2^4$ also guarantees regularity of the phantom crossing point, i.e. $w_{\rm DE}=-1$. Analysis is not straightforward around this point, but there remains sufficient parameter space to ensure that $c_{\rm s}^2 > 0$. In the actual numerical analysis, we check for the ghost and gradient stability by implementing the full expressions from Ref. \cite{DeFelice:2016ucp}, which correctly takes into account the effect of matter. As we will show in Fig. \ref{fig01}, numerical analysis indeed confirms $-1/12 \lesssim \tilde{m}_4^2 / M_p^2 \lesssim 0$ for $0<z<2$.

Furthermore, by specifying the $w_{\rm{DE}}(a)$ through Eq. (\ref{eq:cpl}), one equivalently fixes the background EFT functions $\Lambda$ and $c$. With this setup, it remains to specify the EFT functions $M^4_2$ and $\tilde{m}_4^2$ as functions of scale factor. As kineticity $M^4_2$ is poorly constrained by current observations \cite{deBoe:2024gpf,Ye:2024ywg,Frusciante:2018jzw,Raveri:2019mxg}, without loss of generality we can adopt the phenomenological parametrization \cite{Pan:2025psn}
\begin{equation}
    M_2^4=\sum_{n=0}^{2}c^{M}_n\Omega_{\rm{DE}}^n
\end{equation}
and marginalize over the free coefficients $c^M_{0,1,2}$. The DE fraction is defined as $\Omega_{\rm{DE}}(a)=1-\rho_{\rm{matter}}(a)/3M_p^2H^2(a)$, where $\rho_{\rm{matter}}$ refers to the total energy density of all species except for the DE.

\begin{table}[ht]
\centering
\begin{tabular}{cc}
\hline
Parameter                  & Prior                        \\
\hline
$\Omega_{\rm b} h^2$                 & $\mathcal{U}[0.005,0.1]$       \\
$\Omega_{\rm c} h^2$                 & $\mathcal{U}[0.001,0.99]$       \\
$H_0$ [km/s/Mpc]           & $\mathcal{U}[20,100]$           \\
$n_{\rm s}$                    & $\mathcal{U}[0.8,1.2]$           \\
$\ln(10^{10}A_{\rm s})$          & $\mathcal{U}[1.61,3.91]$          \\
$\tau_{\rm reio}$                   & $\mathcal{U}[0.01,0.8]$         \\
$w_0$                    & $\mathcal{U}[-3,1]$           \\
$w_a$                    & $\mathcal{U}[-3,2]$           \\
\hline
$\tilde{m}^{2}_{4,i}$    & $\mathcal{U}[-1,1]$            \\
$c^{M}_{n}$         & $\mathcal{U}[-10,10]$            \\
\hline
\end{tabular}
\caption{Summary of the uniform priors on the MCMC parameters.}
\label{tab:priors}
\end{table}

To maximize theoretical freedom in the operator of interest $\delta g^{00}R^{(3)}$ and minimize the bias induced by assuming a certain parametrization, we adopt a model-independent reconstruction technique where $\tilde{m}_4^2(a)$ is a free function interpolated over five free nodes $\tilde{m}_{4,i}^2\equiv\tilde{m}_4^2(a_i)$ located at fixed uniformly spaced scale factor points $a_i=[0.4, 0.55, 0.7, 0.85, 1.0]$, corresponding to $z=[1.5, 0.82, 0.43, 0.18, 0]$. We use Gaussian Process interpolation with the standard square exponential kernel function and a correlation length corresponding to the node spacing $l=\Delta a=0.15$. Gaussian Process interpolation has the desired property that the interpolated function is an interpolation between the nodes ($0.4<a<1$) and smoothly extrapolates to the mean value of the nodes outside of the interpolation range ($a<0.4$), ensuring well-behaved asymptotics of $\tilde{m}_4^2$. We start evolving the scalar field perturbations and check for theoretical stability after $a=0.1$. This is appropriate since we are only interested in the low redshift behavior of the theory to stabilize the NEC indicated by DESI \cite{DESI:2025fii} data. Together with the six standard cosmological parameters $\{\Omega_{\rm c}h^2,\Omega_{\rm b}h^2,H_0,A_{\rm s},n_{\rm s},\tau_{\rm reio}\}$, the parametrization coefficients $\{c^{M}_n\}$ and the reconstruction nodes $\{\tilde{m}_{4,i}^2\}$ are sampled in the Monte Carlo Markov chain (MCMC) analysis with uniform priors, summarized in Table.\ref{tab:priors}, to derive the posterior constraints.

We use \texttt{Cobaya} \cite{Torrado:2020dgo,2019ascl.soft10019T} to perform MCMC and \texttt{EFTCAMB} \cite{Hu:2013twa,Raveri:2014cka}, an extension to \texttt{CAMB} \cite{Lewis:1999bs,Howlett:2012mh} to fully implement Eq. (\ref{action01}) and the corresponding ghost and gradient stability checks, to compute cosmology. To have full constraining power on the model we combine DESI DR2 BAO \cite{DESI:2025zgx,DESI:2025zpo} with the full CMB and SNIa observations. For CMB we use the Camspec Planck PR4 high-$\ell$ TTTEEE data \cite{Efstathiou:2019mdh} and Planck PR3 low-$\ell$ TTEE \cite{Planck:2019nip} as well as Planck PR4 lensing \cite{Carron:2022eyg}. For SNIa we use the DES Y5 SNIa compilation \cite{DES:2024jxu}, see also \cite{Efstathiou:2024xcq,Peng:2025nez,Huang:2025som,Gialamas:2024lyw,Park:2024pew,Park:2024vrw,Park:2025azv,Silva:2025hxw,Scherer:2025esj} for some recent discussion on data combination.


{\it Results.---} The main data analysis result is summarized in Fig. \ref{fig01}. For comparison, we further include a $\delta g^{00}R^{(3)}$ only result with $M_2^4=0$. As expected, the left panel of Fig. \ref{fig01} proves that with the ``beyond Horndeski'' operator $\delta g^{00}R^{(3)}$, combined with the kineticity operator $(\delta g^{00})^2$, is able to support the potential NEC violation indicated by the DESI BAO observation. Moreover, the $\delta g^{00}R^{(3)}$ only result proves the theoretical argument made previously that the $\delta g^{00}R^{(3)}$ operator alone cannot stabilize NEC violation (i.e. the red contour hits a hard boundary at the edge of the gray phantom crossing region), but $(\delta g^{00})^2$ is further needed to guarantee ghost stability while $\delta g^{00}R^{(3)}$ removes the gradient instability. Note that the kineticity operator on its own cannot stabilize the observed NEC violation either \cite{Ye:2024ywg}.

The right panel of Fig. \ref{fig01} plots the reconstructed functional shape of ${\tilde{m}_4^2}/{M_p^2}$. It shows that current data can already derive non-trivial constraints on the shape of $\tilde{m}_4^2$. Interestingly, it indicates that $\tilde{m}_4^2\sim {\rm const.}$ is a good fit to observation. Furthermore, the reconstruction yields a $\sim2\sigma$ hint for $\tilde{m}_4^2 \delta g^{00}R^{(3)}\ne0$ under the simplified assumption that only $M_2^4$ and $\tilde{m}_4^2$ are non vanishing in the EFT. As argued previously, $\tilde{m}_4^2 \delta g^{00}R^{(3)}\ne0$ is the requirement of stabilizing phantom crossing. According to \cite{DESI:2025fii}, phantom crossing is commonly recovered in a wide variety of $w_{\rm DE}(a)$ parameterizations as well as non-parametric reconstructions, see also \cite{Ye:2024ywg}. Therefore we expect similar hint of $\tilde{m}_4^2 \delta g^{00}R^{(3)}\ne0$ with other reasonable $w_{\rm DE}(a)$ parameterizations. In Supplemental Material we checked that using alternative SNIa dataset produces statistically consistent results and does not change the conclusion.


{\it Conclusion and discussion.---} In this Letter, we showed that the ``beyond Horndeski'' physics, characterized by the operator $\delta g^{00}R^{(3)}$, which is crucial for fully stabilizing NEC violation in the primordial Universe, also stabilizes NEC violation in DE. Current data shows a $\sim2\sigma$ hint of its existence assuming only the EFT operators $\delta g^{00}R^{(3)}$ and $(\delta g^{00})^2$ are present. Our result establishes the current accelerated expanding Universe as a potential testbed of the ``beyond Horndeski'' physics underlying fully stable NEC violation. It further unveils a possible connection between the primordial Universe and the DE dominated Universe, both featuring an NEC-violating phase prior to accelerated expansion.

We stress that the hint noticed in this letter is only $2\sigma$ and not conclusive. More data in the future is needed to further clarify whether it is physical or statistical fluctuation. Also, the hint is obtained by reconstruction method that introduces five new parameters as interpolation nodes, which will be severely punished if Bayesian model comparison is concerned. We leave detailed statistical analysis and model comparison to future study when more data is available.


Nonetheless, scientific impacts would remain if the hint for the ``beyond Horndeski'' physics was not supported by future data.
Actually, if non-minimal coupling (NMC) $F(\phi)R$ is present, the observation can be explained without invoking ``beyond Horndeski'' physics \cite{Ye:2024ywg,Pan:2025psn, Ye:2024ywg, Ye:2024zpk, Wolf:2024stt}.\footnote{We also note that stable NEC violation with a minimally coupled theory \cite{Ye:2019frg,Ye:2019sth} can be realized within the covariant degenerate higher-order scalar-tensor (DHOST) theory framework \cite{Langlois:2015cwa,BenAchour:2016cay}.}
Within Horndeski theory, $X$-dependent NMC functions have been largely ruled out by observational constraints on the deviation in the GW speed $c_{\rm T}$ from the speed of light, where $X \equiv \nabla_\mu\phi\nabla^\mu\phi$.
However, our work expands the understanding of the feasibility of a phantom crossing of $w_\text{DE}$ by considering ``beyond Horndeski'' physics, which represents an important class of scalar-tensor theories that permit NMC in the form $F(\phi,X)R$ while satisfying $c_{\rm T}=1$ (see e.g., \cite{Creminelli:2017sry,Langlois:2017dyl}). Therefore, ruling out ``beyond Horndeski'' physics through future data would help reduce the dimensionality of the NMC function from two to one, i.e., $F(\phi, X) \rightarrow F(\phi)$, which greatly reduces the degrees of freedom of the theory.\footnote{Strictly speaking, to fully rule out $X$-dependence in the NMC function, one should also examine DHOST theories, which lies beyond the scope of the present work. However, our study represents a meaningful step in this direction.}
This would not only significantly solidify the assumption of a single-variable NMC function, but also effectively enhance the constraining power of observational data.
\\

We thank Yun-Song Piao for valuable discussions.
Y. C. is supported in part by the National Natural Science Foundation of China (Grants No. 12575066, No. 11905224) and the Natural Science Foundation of Henan Province (Grant No. 242300420231).
G. Y. acknowledges support by NWO and the Dutch Ministry of Education, Culture and Science (OCW) (Grant No. VI.Vidi.192.069). Some of the plots were made with \texttt{GetDist} \cite{Lewis:2019xzd}. The authors acknowledge computational support from the ALICE cluster of Leiden University.
The data that support the findings of this article are openly available \cite{zenodo_data}.




\bibliographystyle{utphys}
\bibliography{MyRef}

\providecommand{\href}[2]{#2}\begingroup\raggedright\begin{thebibliography}{100}

\bibitem{Planck:2019nip}
{\bfseries Planck} Collaboration, N.~Aghanim {\em et~al.}, ``{Planck 2018
  results. V. CMB power spectra and likelihoods},''
  \href{http://dx.doi.org/10.1051/0004-6361/201936386}{{\em Astron. Astrophys.}
  {\bfseries 641} (2020) A5}, \href{http://arxiv.org/abs/1907.12875}{{\ttfamily
  arXiv:1907.12875 [astro-ph.CO]}}.

\bibitem{ACT:2025fju}
{\bfseries ACT} Collaboration, T.~Louis {\em et~al.}, ``{The Atacama Cosmology
  Telescope: DR6 Power Spectra, Likelihoods and $\Lambda$CDM Parameters},''
  \href{http://arxiv.org/abs/2503.14452}{{\ttfamily arXiv:2503.14452
  [astro-ph.CO]}}.

\bibitem{Guth:1980zm}
A.~H. Guth, ``{The Inflationary Universe: A Possible Solution to the Horizon
  and Flatness Problems},''
  \href{http://dx.doi.org/10.1103/PhysRevD.23.347}{{\em Phys. Rev. D}
  {\bfseries 23} (1981) 347--356}.

\bibitem{Starobinsky:1980te}
A.~A. Starobinsky, ``{A New Type of Isotropic Cosmological Models Without
  Singularity},'' \href{http://dx.doi.org/10.1016/0370-2693(80)90670-X}{{\em
  Phys. Lett. B} {\bfseries 91} (1980) 99--102}.

\bibitem{Linde:1981mu}
A.~D. Linde, ``{A New Inflationary Universe Scenario: A Possible Solution of
  the Horizon, Flatness, Homogeneity, Isotropy and Primordial Monopole
  Problems},'' \href{http://dx.doi.org/10.1016/0370-2693(82)91219-9}{{\em Phys.
  Lett. B} {\bfseries 108} (1982) 389--393}.

\bibitem{Albrecht:1982wi}
A.~Albrecht and P.~J. Steinhardt, ``{Cosmology for Grand Unified Theories with
  Radiatively Induced Symmetry Breaking},''
  \href{http://dx.doi.org/10.1103/PhysRevLett.48.1220}{{\em Phys. Rev. Lett.}
  {\bfseries 48} (1982) 1220--1223}.

\bibitem{DESI:2024lzq}
{\bfseries DESI} Collaboration, A.~G. Adame {\em et~al.}, ``{DESI 2024 IV:
  Baryon Acoustic Oscillations from the Lyman alpha forest},''
  \href{http://dx.doi.org/10.1088/1475-7516/2025/01/124}{{\em JCAP} {\bfseries
  01} (2025) 124}, \href{http://arxiv.org/abs/2404.03001}{{\ttfamily
  arXiv:2404.03001 [astro-ph.CO]}}.

\bibitem{DESI:2024uvr}
{\bfseries DESI} Collaboration, A.~G. Adame {\em et~al.}, ``{DESI 2024 III:
  baryon acoustic oscillations from galaxies and quasars},''
  \href{http://dx.doi.org/10.1088/1475-7516/2025/04/012}{{\em JCAP} {\bfseries
  04} (2025) 012}, \href{http://arxiv.org/abs/2404.03000}{{\ttfamily
  arXiv:2404.03000 [astro-ph.CO]}}.

\bibitem{DESI:2024mwx}
{\bfseries DESI} Collaboration, A.~G. Adame {\em et~al.}, ``{DESI 2024 VI:
  cosmological constraints from the measurements of baryon acoustic
  oscillations},'' \href{http://dx.doi.org/10.1088/1475-7516/2025/02/021}{{\em
  JCAP} {\bfseries 02} (2025) 021},
  \href{http://arxiv.org/abs/2404.03002}{{\ttfamily arXiv:2404.03002
  [astro-ph.CO]}}.

\bibitem{DESI:2025zgx}
{\bfseries DESI} Collaboration, M.~Abdul~Karim {\em et~al.}, ``{DESI DR2
  Results II: Measurements of Baryon Acoustic Oscillations and Cosmological
  Constraints},'' \href{http://arxiv.org/abs/2503.14738}{{\ttfamily
  arXiv:2503.14738 [astro-ph.CO]}}.

\bibitem{DESI:2025zpo}
M.~A. Karim {\em et~al.}, ``{DESI DR2 Results I: Baryon Acoustic Oscillations
  from the Lyman Alpha Forest},''
  \href{http://dx.doi.org/10.48550/arXiv.2503.14739}{{\em arXiv e-prints}
  (Mar., 2025) arXiv:2503.14739},
  \href{http://arxiv.org/abs/2503.14739}{{\ttfamily arXiv:2503.14739
  [astro-ph.CO]}}.

\bibitem{Brandenberger:2025hof}
R.~Brandenberger, ``{Why the DESI Results Should Not Be A Surprise},''
  \href{http://arxiv.org/abs/2503.17659}{{\ttfamily arXiv:2503.17659
  [astro-ph.CO]}}.

\bibitem{Borde:2001nh}
A.~Borde, A.~H. Guth, and A.~Vilenkin, ``{Inflationary space-times are
  incompletein past directions},''
  \href{http://dx.doi.org/10.1103/PhysRevLett.90.151301}{{\em Phys. Rev. Lett.}
  {\bfseries 90} (2003) 151301},
  \href{http://arxiv.org/abs/gr-qc/0110012}{{\ttfamily arXiv:gr-qc/0110012}}.

\bibitem{Penrose:1964wq}
R.~Penrose, ``{Gravitational collapse and space-time singularities},''
  \href{http://dx.doi.org/10.1103/PhysRevLett.14.57}{{\em Phys. Rev. Lett.}
  {\bfseries 14} (1965) 57--59}.

\bibitem{Rubakov:2014jja}
V.~A. Rubakov, ``{The Null Energy Condition and its violation},''
  \href{http://dx.doi.org/10.3367/UFNe.0184.201402b.0137}{{\em Phys. Usp.}
  {\bfseries 57} (2014) 128--142},
  \href{http://arxiv.org/abs/1401.4024}{{\ttfamily arXiv:1401.4024 [hep-th]}}.

\bibitem{DESI:2025fii}
{\bfseries DESI} Collaboration, K.~Lodha {\em et~al.}, ``{Extended Dark Energy
  analysis using DESI DR2 BAO measurements},''
  \href{http://arxiv.org/abs/2503.14743}{{\ttfamily arXiv:2503.14743
  [astro-ph.CO]}}.

\bibitem{Feng:2004ad}
B.~Feng, X.-L. Wang, and X.-M. Zhang, ``{Dark energy constraints from the
  cosmic age and supernova},''
  \href{http://dx.doi.org/10.1016/j.physletb.2004.12.071}{{\em Phys. Lett. B}
  {\bfseries 607} (2005) 35--41},
  \href{http://arxiv.org/abs/astro-ph/0404224}{{\ttfamily
  arXiv:astro-ph/0404224}}.

\bibitem{Wei:2005nw}
H.~Wei, R.-G. Cai, and D.-F. Zeng, ``{Hessence: A New view of quintom dark
  energy},'' \href{http://dx.doi.org/10.1088/0264-9381/22/16/005}{{\em Class.
  Quant. Grav.} {\bfseries 22} (2005) 3189--3202},
  \href{http://arxiv.org/abs/hep-th/0501160}{{\ttfamily arXiv:hep-th/0501160}}.

\bibitem{Caldwell:2005ai}
R.~R. Caldwell and M.~Doran, ``{Dark-energy evolution across the
  cosmological-constant boundary},''
  \href{http://dx.doi.org/10.1103/PhysRevD.72.043527}{{\em Phys. Rev. D}
  {\bfseries 72} (2005) 043527},
  \href{http://arxiv.org/abs/astro-ph/0501104}{{\ttfamily
  arXiv:astro-ph/0501104}}.

\bibitem{Cai:2007zv}
Y.-F. Cai, T.~Qiu, R.~Brandenberger, Y.-S. Piao, and X.~Zhang, ``{On
  Perturbations of Quintom Bounce},''
  \href{http://dx.doi.org/10.1088/1475-7516/2008/03/013}{{\em JCAP} {\bfseries
  03} (2008) 013}, \href{http://arxiv.org/abs/0711.2187}{{\ttfamily
  arXiv:0711.2187 [hep-th]}}.

\bibitem{Horndeski:1974wa}
G.~W. Horndeski, ``{Second-order scalar-tensor field equations in a
  four-dimensional space},'' \href{http://dx.doi.org/10.1007/BF01807638}{{\em
  Int. J. Theor. Phys.} {\bfseries 10} (1974) 363--384}.

\bibitem{Creminelli:2006xe}
P.~Creminelli, M.~A. Luty, A.~Nicolis, and L.~Senatore, ``{Starting the
  Universe: Stable Violation of the Null Energy Condition and Non-standard
  Cosmologies},'' \href{http://dx.doi.org/10.1088/1126-6708/2006/12/080}{{\em
  JHEP} {\bfseries 12} (2006) 080},
  \href{http://arxiv.org/abs/hep-th/0606090}{{\ttfamily arXiv:hep-th/0606090}}.

\bibitem{Cheung:2007st}
C.~Cheung, P.~Creminelli, A.~L. Fitzpatrick, J.~Kaplan, and L.~Senatore, ``{The
  Effective Field Theory of Inflation},''
  \href{http://dx.doi.org/10.1088/1126-6708/2008/03/014}{{\em JHEP} {\bfseries
  03} (2008) 014}, \href{http://arxiv.org/abs/0709.0293}{{\ttfamily
  arXiv:0709.0293 [hep-th]}}.

\bibitem{Weinberg:2008hq}
S.~Weinberg, ``{Effective Field Theory for Inflation},''
  \href{http://dx.doi.org/10.1103/PhysRevD.77.123541}{{\em Phys. Rev. D}
  {\bfseries 77} (2008) 123541},
  \href{http://arxiv.org/abs/0804.4291}{{\ttfamily arXiv:0804.4291 [hep-th]}}.

\bibitem{Gubitosi:2012hu}
G.~Gubitosi, F.~Piazza, and F.~Vernizzi, ``{The Effective Field Theory of Dark
  Energy},'' \href{http://dx.doi.org/10.1088/1475-7516/2013/02/032}{{\em JCAP}
  {\bfseries 02} (2013) 032}, \href{http://arxiv.org/abs/1210.0201}{{\ttfamily
  arXiv:1210.0201 [hep-th]}}.

\bibitem{Bloomfield:2012ff}
J.~K. Bloomfield, E.~E. Flanagan, M.~Park, and S.~Watson, ``{Dark energy or
  modified gravity? An effective field theory approach},''
  \href{http://dx.doi.org/10.1088/1475-7516/2013/08/010}{{\em JCAP} {\bfseries
  08} (2013) 010}, \href{http://arxiv.org/abs/1211.7054}{{\ttfamily
  arXiv:1211.7054 [astro-ph.CO]}}.

\bibitem{Gleyzes:2013ooa}
J.~Gleyzes, D.~Langlois, F.~Piazza, and F.~Vernizzi, ``{Essential Building
  Blocks of Dark Energy},''
  \href{http://dx.doi.org/10.1088/1475-7516/2013/08/025}{{\em JCAP} {\bfseries
  08} (2013) 025}, \href{http://arxiv.org/abs/1304.4840}{{\ttfamily
  arXiv:1304.4840 [hep-th]}}.

\bibitem{Carroll:2003st}
S.~M. Carroll, M.~Hoffman, and M.~Trodden, ``{Can the dark energy
  equation-of-state parameter $w$ be less than $-1$?},''
  \href{http://dx.doi.org/10.1103/PhysRevD.68.023509}{{\em Phys. Rev. D}
  {\bfseries 68} (2003) 023509},
  \href{http://arxiv.org/abs/astro-ph/0301273}{{\ttfamily
  arXiv:astro-ph/0301273}}.

\bibitem{Cline:2003gs}
J.~M. Cline, S.~Jeon, and G.~D. Moore, ``{The Phantom menaced: Constraints on
  low-energy effective ghosts},''
  \href{http://dx.doi.org/10.1103/PhysRevD.70.043543}{{\em Phys. Rev. D}
  {\bfseries 70} (2004) 043543},
  \href{http://arxiv.org/abs/hep-ph/0311312}{{\ttfamily arXiv:hep-ph/0311312}}.

\bibitem{Dubovsky:2005xd}
S.~Dubovsky, T.~Gregoire, A.~Nicolis, and R.~Rattazzi, ``{Null energy condition
  and superluminal propagation},''
  \href{http://dx.doi.org/10.1088/1126-6708/2006/03/025}{{\em JHEP} {\bfseries
  03} (2006) 025}, \href{http://arxiv.org/abs/hep-th/0512260}{{\ttfamily
  arXiv:hep-th/0512260}}.

\bibitem{Nicolis:2009qm}
A.~Nicolis, R.~Rattazzi, and E.~Trincherini, ``{Energy's and amplitudes'
  positivity},'' \href{http://dx.doi.org/10.1007/JHEP05(2010)095}{{\em JHEP}
  {\bfseries 05} (2010) 095}, \href{http://arxiv.org/abs/0912.4258}{{\ttfamily
  arXiv:0912.4258 [hep-th]}}. [Erratum: JHEP 11, 128 (2011)].

\bibitem{Creminelli:2010ba}
P.~Creminelli, A.~Nicolis, and E.~Trincherini, ``{Galilean Genesis: An
  Alternative to inflation},''
  \href{http://dx.doi.org/10.1088/1475-7516/2010/11/021}{{\em JCAP} {\bfseries
  11} (2010) 021}, \href{http://arxiv.org/abs/1007.0027}{{\ttfamily
  arXiv:1007.0027 [hep-th]}}.

\bibitem{Creminelli:2012my}
P.~Creminelli, K.~Hinterbichler, J.~Khoury, A.~Nicolis, and E.~Trincherini,
  ``{Subluminal Galilean Genesis},''
  \href{http://dx.doi.org/10.1007/JHEP02(2013)006}{{\em JHEP} {\bfseries 02}
  (2013) 006}, \href{http://arxiv.org/abs/1209.3768}{{\ttfamily arXiv:1209.3768
  [hep-th]}}.

\bibitem{Moghtaderi:2025cns}
E.~Moghtaderi, B.~R. Hull, J.~Quintin, and G.~Geshnizjani, ``{How Much NEC
  Breaking Can the Universe Endure?},''
  \href{http://arxiv.org/abs/2503.19955}{{\ttfamily arXiv:2503.19955 [gr-qc]}}.

\bibitem{Easson:2011zy}
D.~A. Easson, I.~Sawicki, and A.~Vikman, ``{G-Bounce},''
  \href{http://dx.doi.org/10.1088/1475-7516/2011/11/021}{{\em JCAP} {\bfseries
  11} (2011) 021}, \href{http://arxiv.org/abs/1109.1047}{{\ttfamily
  arXiv:1109.1047 [hep-th]}}.

\bibitem{Battarra:2014tga}
L.~Battarra, M.~Koehn, J.-L. Lehners, and B.~A. Ovrut, ``{Cosmological
  Perturbations Through a Non-Singular Ghost-Condensate/Galileon Bounce},''
  \href{http://dx.doi.org/10.1088/1475-7516/2014/07/007}{{\em JCAP} {\bfseries
  07} (2014) 007}, \href{http://arxiv.org/abs/1404.5067}{{\ttfamily
  arXiv:1404.5067 [hep-th]}}.

\bibitem{Koehn:2015vvy}
M.~Koehn, J.-L. Lehners, and B.~Ovrut, ``{Nonsingular bouncing cosmology:
  Consistency of the effective description},''
  \href{http://dx.doi.org/10.1103/PhysRevD.93.103501}{{\em Phys. Rev. D}
  {\bfseries 93} no.~10, (2016) 103501},
  \href{http://arxiv.org/abs/1512.03807}{{\ttfamily arXiv:1512.03807
  [hep-th]}}.

\bibitem{Qiu:2015nha}
T.~Qiu and Y.-T. Wang, ``{G-Bounce Inflation: Towards Nonsingular Inflation
  Cosmology with Galileon Field},''
  \href{http://dx.doi.org/10.1007/JHEP04(2015)130}{{\em JHEP} {\bfseries 04}
  (2015) 130}, \href{http://arxiv.org/abs/1501.03568}{{\ttfamily
  arXiv:1501.03568 [astro-ph.CO]}}.

\bibitem{Ijjas:2016tpn}
A.~Ijjas and P.~J. Steinhardt, ``{Classically stable nonsingular cosmological
  bounces},'' \href{http://dx.doi.org/10.1103/PhysRevLett.117.121304}{{\em
  Phys. Rev. Lett.} {\bfseries 117} no.~12, (2016) 121304},
  \href{http://arxiv.org/abs/1606.08880}{{\ttfamily arXiv:1606.08880 [gr-qc]}}.

\bibitem{Ijjas:2016vtq}
A.~Ijjas and P.~J. Steinhardt, ``{Fully stable cosmological solutions with a
  non-singular classical bounce},''
  \href{http://dx.doi.org/10.1016/j.physletb.2016.11.047}{{\em Phys. Lett. B}
  {\bfseries 764} (2017) 289--294},
  \href{http://arxiv.org/abs/1609.01253}{{\ttfamily arXiv:1609.01253 [gr-qc]}}.

\bibitem{Dobre:2017pnt}
D.~A. Dobre, A.~V. Frolov, J.~T. G\'alvez~Ghersi, S.~Ramazanov, and A.~Vikman,
  ``{Unbraiding the Bounce: Superluminality around the Corner},''
  \href{http://dx.doi.org/10.1088/1475-7516/2018/03/020}{{\em JCAP} {\bfseries
  03} (2018) 020}, \href{http://arxiv.org/abs/1712.10272}{{\ttfamily
  arXiv:1712.10272 [gr-qc]}}.

\bibitem{Libanov:2016kfc}
M.~Libanov, S.~Mironov, and V.~Rubakov, ``{Generalized Galileons: instabilities
  of bouncing and Genesis cosmologies and modified Genesis},''
  \href{http://dx.doi.org/10.1088/1475-7516/2016/08/037}{{\em JCAP} {\bfseries
  08} (2016) 037}, \href{http://arxiv.org/abs/1605.05992}{{\ttfamily
  arXiv:1605.05992 [hep-th]}}.

\bibitem{Kobayashi:2016xpl}
T.~Kobayashi, ``{Generic instabilities of nonsingular cosmologies in Horndeski
  theory: A no-go theorem},''
  \href{http://dx.doi.org/10.1103/PhysRevD.94.043511}{{\em Phys. Rev. D}
  {\bfseries 94} no.~4, (2016) 043511},
  \href{http://arxiv.org/abs/1606.05831}{{\ttfamily arXiv:1606.05831
  [hep-th]}}.

\bibitem{Deffayet:2011gz}
C.~Deffayet, X.~Gao, D.~A. Steer, and G.~Zahariade, ``{From k-essence to
  generalised Galileons},''
  \href{http://dx.doi.org/10.1103/PhysRevD.84.064039}{{\em Phys. Rev. D}
  {\bfseries 84} (2011) 064039},
  \href{http://arxiv.org/abs/1103.3260}{{\ttfamily arXiv:1103.3260 [hep-th]}}.

\bibitem{Kobayashi:2011nu}
T.~Kobayashi, M.~Yamaguchi, and J.~Yokoyama, ``{Generalized G-inflation:
  Inflation with the most general second-order field equations},''
  \href{http://dx.doi.org/10.1143/PTP.126.511}{{\em Prog. Theor. Phys.}
  {\bfseries 126} (2011) 511--529},
  \href{http://arxiv.org/abs/1105.5723}{{\ttfamily arXiv:1105.5723 [hep-th]}}.

\bibitem{Cai:2016thi}
Y.~Cai, Y.~Wan, H.-G. Li, T.~Qiu, and Y.-S. Piao, ``{The Effective Field Theory
  of nonsingular cosmology},''
  \href{http://dx.doi.org/10.1007/JHEP01(2017)090}{{\em JHEP} {\bfseries 01}
  (2017) 090}, \href{http://arxiv.org/abs/1610.03400}{{\ttfamily
  arXiv:1610.03400 [gr-qc]}}.

\bibitem{Creminelli:2016zwa}
P.~Creminelli, D.~Pirtskhalava, L.~Santoni, and E.~Trincherini, ``{Stability of
  Geodesically Complete Cosmologies},''
  \href{http://dx.doi.org/10.1088/1475-7516/2016/11/047}{{\em JCAP} {\bfseries
  11} (2016) 047}, \href{http://arxiv.org/abs/1610.04207}{{\ttfamily
  arXiv:1610.04207 [hep-th]}}.

\bibitem{Cai:2017dyi}
Y.~Cai and Y.-S. Piao, ``{A covariant Lagrangian for stable nonsingular
  bounce},'' \href{http://dx.doi.org/10.1007/JHEP09(2017)027}{{\em JHEP}
  {\bfseries 09} (2017) 027}, \href{http://arxiv.org/abs/1705.03401}{{\ttfamily
  arXiv:1705.03401 [gr-qc]}}.

\bibitem{Kolevatov:2017voe}
R.~Kolevatov, S.~Mironov, N.~Sukhov, and V.~Volkova, ``{Cosmological bounce and
  Genesis beyond Horndeski},''
  \href{http://dx.doi.org/10.1088/1475-7516/2017/08/038}{{\em JCAP} {\bfseries
  08} (2017) 038}, \href{http://arxiv.org/abs/1705.06626}{{\ttfamily
  arXiv:1705.06626 [hep-th]}}.

\bibitem{Cai:2017dxl}
Y.~Cai and Y.-S. Piao, ``{Higher order derivative coupling to gravity and its
  cosmological implications},''
  \href{http://dx.doi.org/10.1103/PhysRevD.96.124028}{{\em Phys. Rev. D}
  {\bfseries 96} no.~12, (2017) 124028},
  \href{http://arxiv.org/abs/1707.01017}{{\ttfamily arXiv:1707.01017 [gr-qc]}}.

\bibitem{Cai:2017pga}
Y.~Cai, Y.-T. Wang, J.-Y. Zhao, and Y.-S. Piao, ``{Primordial perturbations
  with pre-inflationary bounce},''
  \href{http://dx.doi.org/10.1103/PhysRevD.97.103535}{{\em Phys. Rev. D}
  {\bfseries 97} no.~10, (2018) 103535},
  \href{http://arxiv.org/abs/1709.07464}{{\ttfamily arXiv:1709.07464
  [astro-ph.CO]}}.

\bibitem{Mironov:2018oec}
S.~Mironov, V.~Rubakov, and V.~Volkova, ``{Bounce beyond Horndeski with GR
  asymptotics and $\gamma$-crossing},''
  \href{http://dx.doi.org/10.1088/1475-7516/2018/10/050}{{\em JCAP} {\bfseries
  10} (2018) 050}, \href{http://arxiv.org/abs/1807.08361}{{\ttfamily
  arXiv:1807.08361 [hep-th]}}.

\bibitem{Ye:2019frg}
G.~Ye and Y.-S. Piao, ``{Implication of GW170817 for cosmological bounces},''
  \href{http://dx.doi.org/10.1088/0253-6102/71/4/427}{{\em Commun. Theor.
  Phys.} {\bfseries 71} no.~4, (2019) 427},
  \href{http://arxiv.org/abs/1901.02202}{{\ttfamily arXiv:1901.02202 [gr-qc]}}.

\bibitem{Ye:2019sth}
G.~Ye and Y.-S. Piao, ``{Bounce in general relativity and higher-order
  derivative operators},''
  \href{http://dx.doi.org/10.1103/PhysRevD.99.084019}{{\em Phys. Rev. D}
  {\bfseries 99} no.~8, (2019) 084019},
  \href{http://arxiv.org/abs/1901.08283}{{\ttfamily arXiv:1901.08283 [gr-qc]}}.

\bibitem{Mironov:2019qjt}
S.~Mironov, V.~Rubakov, and V.~Volkova, ``{Genesis with general relativity
  asymptotics in beyond Horndeski theory},''
  \href{http://dx.doi.org/10.1103/PhysRevD.100.083521}{{\em Phys. Rev. D}
  {\bfseries 100} no.~8, (2019) 083521},
  \href{http://arxiv.org/abs/1905.06249}{{\ttfamily arXiv:1905.06249
  [hep-th]}}.

\bibitem{Akama:2019qeh}
S.~Akama, S.~Hirano, and T.~Kobayashi, ``{Primordial non-Gaussianities of
  scalar and tensor perturbations in general bounce cosmology: Evading the
  no-go theorem},'' \href{http://dx.doi.org/10.1103/PhysRevD.101.043529}{{\em
  Phys. Rev. D} {\bfseries 101} no.~4, (2020) 043529},
  \href{http://arxiv.org/abs/1908.10663}{{\ttfamily arXiv:1908.10663 [gr-qc]}}.

\bibitem{Mironov:2019mye}
S.~Mironov, V.~Rubakov, and V.~Volkova, ``{Subluminal cosmological bounce
  beyond Horndeski},''
  \href{http://dx.doi.org/10.1088/1475-7516/2020/05/024}{{\em JCAP} {\bfseries
  05} (2020) 024}, \href{http://arxiv.org/abs/1910.07019}{{\ttfamily
  arXiv:1910.07019 [hep-th]}}.

\bibitem{Ilyas:2020qja}
A.~Ilyas, M.~Zhu, Y.~Zheng, Y.-F. Cai, and E.~N. Saridakis, ``{DHOST Bounce},''
  \href{http://dx.doi.org/10.1088/1475-7516/2020/09/002}{{\em JCAP} {\bfseries
  09} (2020) 002}, \href{http://arxiv.org/abs/2002.08269}{{\ttfamily
  arXiv:2002.08269 [gr-qc]}}.

\bibitem{Ilyas:2020zcb}
A.~Ilyas, M.~Zhu, Y.~Zheng, and Y.-F. Cai, ``{Emergent Universe and Genesis
  from the DHOST Cosmology},''
  \href{http://dx.doi.org/10.1007/JHEP01(2021)141}{{\em JHEP} {\bfseries 01}
  (2021) 141}, \href{http://arxiv.org/abs/2009.10351}{{\ttfamily
  arXiv:2009.10351 [gr-qc]}}.

\bibitem{Zhu:2021whu}
M.~Zhu, A.~Ilyas, Y.~Zheng, Y.-F. Cai, and E.~N. Saridakis, ``{Scalar and
  tensor perturbations in DHOST bounce cosmology},''
  \href{http://dx.doi.org/10.1088/1475-7516/2021/11/045}{{\em JCAP} {\bfseries
  11} no.~11, (2021) 045}, \href{http://arxiv.org/abs/2108.01339}{{\ttfamily
  arXiv:2108.01339 [gr-qc]}}.

\bibitem{Zhu:2021ggm}
M.~Zhu and Y.~Zheng, ``{Improved DHOST Genesis},''
  \href{http://dx.doi.org/10.1007/JHEP11(2021)163}{{\em JHEP} {\bfseries 11}
  (2021) 163}, \href{http://arxiv.org/abs/2109.05277}{{\ttfamily
  arXiv:2109.05277 [gr-qc]}}.

\bibitem{Cai:2022ori}
Y.~Cai, J.~Xu, S.~Zhao, and S.~Zhou, ``{Perturbative unitarity and NEC
  violation in genesis cosmology},''
  \href{http://dx.doi.org/10.1007/JHEP10(2022)140}{{\em JHEP} {\bfseries 10}
  (2022) 140}, \href{http://arxiv.org/abs/2207.11772}{{\ttfamily
  arXiv:2207.11772 [gr-qc]}}. [Erratum: JHEP 11, 063 (2022)].

\bibitem{Akama:2025ows}
S.~Akama, ``{Primordial full bispectra from the general bounce cosmology},''
  \href{http://arxiv.org/abs/2502.14850}{{\ttfamily arXiv:2502.14850
  [astro-ph.CO]}}.

\bibitem{Dehghani:2025udv}
A.~Dehghani, G.~Geshnizjani, and J.~Quintin, ``{Cuscuton Bounce Beyond the
  Linear Regime: Bispectrum and Strong Coupling Constraints},''
  \href{http://arxiv.org/abs/2503.01992}{{\ttfamily arXiv:2503.01992
  [hep-th]}}.

\bibitem{Gleyzes:2014dya}
J.~Gleyzes, D.~Langlois, F.~Piazza, and F.~Vernizzi, ``{Healthy theories beyond
  Horndeski},'' \href{http://dx.doi.org/10.1103/PhysRevLett.114.211101}{{\em
  Phys. Rev. Lett.} {\bfseries 114} no.~21, (2015) 211101},
  \href{http://arxiv.org/abs/1404.6495}{{\ttfamily arXiv:1404.6495 [hep-th]}}.

\bibitem{Gleyzes:2014qga}
J.~Gleyzes, D.~Langlois, F.~Piazza, and F.~Vernizzi, ``{Exploring gravitational
  theories beyond Horndeski},''
  \href{http://dx.doi.org/10.1088/1475-7516/2015/02/018}{{\em JCAP} {\bfseries
  02} (2015) 018}, \href{http://arxiv.org/abs/1408.1952}{{\ttfamily
  arXiv:1408.1952 [astro-ph.CO]}}.

\bibitem{Cai:2020qpu}
Y.~Cai and Y.-S. Piao, ``{Intermittent null energy condition violations during
  inflation and primordial gravitational waves},''
  \href{http://dx.doi.org/10.1103/PhysRevD.103.083521}{{\em Phys. Rev. D}
  {\bfseries 103} no.~8, (2021) 083521},
  \href{http://arxiv.org/abs/2012.11304}{{\ttfamily arXiv:2012.11304 [gr-qc]}}.

\bibitem{Cai:2022nqv}
Y.~Cai and Y.-S. Piao, ``{Generating enhanced primordial GWs during inflation
  with intermittent violation of NEC and diminishment of GW propagating
  speed},'' \href{http://dx.doi.org/10.1007/JHEP06(2022)067}{{\em JHEP}
  {\bfseries 06} (2022) 067}, \href{http://arxiv.org/abs/2201.04552}{{\ttfamily
  arXiv:2201.04552 [gr-qc]}}.

\bibitem{Jiang:2023gfe}
J.-Q. Jiang, Y.~Cai, G.~Ye, and Y.-S. Piao, ``{Broken blue-tilted inflationary
  gravitational waves: a joint analysis of NANOGrav 15-year and BICEP/Keck 2018
  data},'' \href{http://dx.doi.org/10.1088/1475-7516/2024/05/004}{{\em JCAP}
  {\bfseries 05} (2024) 004}, \href{http://arxiv.org/abs/2307.15547}{{\ttfamily
  arXiv:2307.15547 [astro-ph.CO]}}.

\bibitem{Ye:2023tpz}
G.~Ye, M.~Zhu, and Y.~Cai, ``{Null energy condition violation during inflation
  and pulsar timing array observations},''
  \href{http://dx.doi.org/10.1007/JHEP02(2024)008}{{\em JHEP} {\bfseries 02}
  (2024) 008}, \href{http://arxiv.org/abs/2312.10685}{{\ttfamily
  arXiv:2312.10685 [gr-qc]}}.

\bibitem{Pan:2024ydt}
S.~Pan, Y.~Cai, and Y.-S. Piao, ``{Climbing over the potential barrier during
  inflation via null energy condition violation},''
  \href{http://dx.doi.org/10.1140/epjc/s10052-024-13340-1}{{\em Eur. Phys. J.
  C} {\bfseries 84} no.~9, (2024) 976},
  \href{http://arxiv.org/abs/2404.12655}{{\ttfamily arXiv:2404.12655
  [astro-ph.CO]}}.

\bibitem{Cai:2022lec}
Y.~Cai, ``{Generating enhanced parity-violating gravitational waves during
  inflation with violation of the null energy condition},''
  \href{http://dx.doi.org/10.1103/PhysRevD.107.063512}{{\em Phys. Rev. D}
  {\bfseries 107} no.~6, (2023) 063512},
  \href{http://arxiv.org/abs/2212.10893}{{\ttfamily arXiv:2212.10893 [gr-qc]}}.

\bibitem{Jiang:2024woi}
Z.-W. Jiang, Y.~Cai, F.~Wang, and Y.-S. Piao, ``{Parity-violating primordial
  gravitational waves from null energy condition violation},''
  \href{http://dx.doi.org/10.1007/JHEP09(2024)067}{{\em JHEP} {\bfseries 09}
  (2024) 067}, \href{http://arxiv.org/abs/2406.16549}{{\ttfamily
  arXiv:2406.16549 [astro-ph.CO]}}.

\bibitem{Cai:2023uhc}
Y.~Cai, M.~Zhu, and Y.-S. Piao, ``{Primordial Black Holes from Null Energy
  Condition Violation during Inflation},''
  \href{http://dx.doi.org/10.1103/PhysRevLett.133.021001}{{\em Phys. Rev.
  Lett.} {\bfseries 133} no.~2, (2024) 021001},
  \href{http://arxiv.org/abs/2305.10933}{{\ttfamily arXiv:2305.10933 [gr-qc]}}.

\bibitem{Ageeva:2018lko}
Y.~A. Ageeva, O.~A. Evseev, O.~I. Melichev, and V.~A. Rubakov, ``{Horndeski
  Genesis: strong coupling and absence thereof},''
  \href{http://dx.doi.org/10.1051/epjconf/201819107010}{{\em EPJ Web Conf.}
  {\bfseries 191} (2018) 07010},
  \href{http://arxiv.org/abs/1810.00465}{{\ttfamily arXiv:1810.00465
  [hep-th]}}.

\bibitem{Ageeva:2020gti}
Y.~Ageeva, O.~Evseev, O.~Melichev, and V.~Rubakov, ``{Toward evading the strong
  coupling problem in Horndeski genesis},''
  \href{http://dx.doi.org/10.1103/PhysRevD.102.023519}{{\em Phys. Rev. D}
  {\bfseries 102} no.~2, (2020) 023519},
  \href{http://arxiv.org/abs/2003.01202}{{\ttfamily arXiv:2003.01202
  [hep-th]}}.

\bibitem{Ageeva:2021yik}
Y.~Ageeva, P.~Petrov, and V.~Rubakov, ``{Nonsingular cosmological models with
  strong gravity in the past},''
  \href{http://dx.doi.org/10.1103/PhysRevD.104.063530}{{\em Phys. Rev. D}
  {\bfseries 104} no.~6, (2021) 063530},
  \href{http://arxiv.org/abs/2104.13412}{{\ttfamily arXiv:2104.13412
  [hep-th]}}.

\bibitem{GilChoi:2025hbs}
H.~Gil~Choi, P.~Petrov, and M.~Yamaguchi, ``{Can Horndeski Genesis be
  Nonpathological?},'' \href{http://arxiv.org/abs/2503.02626}{{\ttfamily
  arXiv:2503.02626 [hep-th]}}.

\bibitem{Chevallier:2000qy}
M.~Chevallier and D.~Polarski, ``{Accelerating universes with scaling dark
  matter},'' \href{http://dx.doi.org/10.1142/S0218271801000822}{{\em Int. J.
  Mod. Phys. D} {\bfseries 10} (2001) 213--224},
  \href{http://arxiv.org/abs/gr-qc/0009008}{{\ttfamily arXiv:gr-qc/0009008}}.

\bibitem{Linder:2002et}
E.~V. Linder, ``{Exploring the expansion history of the universe},''
  \href{http://dx.doi.org/10.1103/PhysRevLett.90.091301}{{\em Phys. Rev. Lett.}
  {\bfseries 90} (2003) 091301},
  \href{http://arxiv.org/abs/astro-ph/0208512}{{\ttfamily
  arXiv:astro-ph/0208512}}.

\bibitem{deBoe:2024gpf}
D.~de~Boe, G.~Ye, F.~Renzi, I.~S. Albuquerque, N.~Frusciante, and A.~Silvestri,
  ``{Phenomenology of Horndeski gravity under positivity bounds},''
  \href{http://dx.doi.org/10.1088/1475-7516/2024/08/029}{{\em JCAP} {\bfseries
  08} (2024) 029}, \href{http://arxiv.org/abs/2403.13096}{{\ttfamily
  arXiv:2403.13096 [astro-ph.CO]}}.

\bibitem{Ye:2024ywg}
G.~Ye, M.~Martinelli, B.~Hu, and A.~Silvestri, ``{Non-minimally coupled gravity
  as a physically viable fit to DESI 2024 BAO},''
  \href{http://arxiv.org/abs/2407.15832}{{\ttfamily arXiv:2407.15832
  [astro-ph.CO]}}.

\bibitem{Frusciante:2018jzw}
N.~Frusciante, S.~Peirone, S.~Casas, and N.~A. Lima, ``{Cosmology of surviving
  Horndeski theory: The road ahead},''
  \href{http://dx.doi.org/10.1103/PhysRevD.99.063538}{{\em Phys. Rev. D}
  {\bfseries 99} no.~6, (2019) 063538},
  \href{http://arxiv.org/abs/1810.10521}{{\ttfamily arXiv:1810.10521
  [astro-ph.CO]}}.

\bibitem{Raveri:2019mxg}
M.~Raveri, ``{Reconstructing Gravity on Cosmological Scales},''
  \href{http://dx.doi.org/10.1103/PhysRevD.101.083524}{{\em Phys. Rev. D}
  {\bfseries 101} no.~8, (2020) 083524},
  \href{http://arxiv.org/abs/1902.01366}{{\ttfamily arXiv:1902.01366
  [astro-ph.CO]}}.

\bibitem{Kase:2014yya}
R.~Kase and S.~Tsujikawa, ``{Cosmology in generalized Horndeski theories with
  second-order equations of motion},''
  \href{http://dx.doi.org/10.1103/PhysRevD.90.044073}{{\em Phys. Rev. D}
  {\bfseries 90} (2014) 044073},
  \href{http://arxiv.org/abs/1407.0794}{{\ttfamily arXiv:1407.0794 [hep-th]}}.

\bibitem{DeFelice:2016ucp}
A.~De~Felice, N.~Frusciante, and G.~Papadomanolakis, ``{On the stability
  conditions for theories of modified gravity in the presence of matter
  fields},'' \href{http://dx.doi.org/10.1088/1475-7516/2017/03/027}{{\em JCAP}
  {\bfseries 03} (2017) 027}, \href{http://arxiv.org/abs/1609.03599}{{\ttfamily
  arXiv:1609.03599 [gr-qc]}}.

\bibitem{Frusciante:2016xoj}
N.~Frusciante, G.~Papadomanolakis, and A.~Silvestri, ``{An Extended action for
  the effective field theory of dark energy: a stability analysis and a
  complete guide to the mapping at the basis of EFTCAMB},''
  \href{http://dx.doi.org/10.1088/1475-7516/2016/07/018}{{\em JCAP} {\bfseries
  07} (2016) 018}, \href{http://arxiv.org/abs/1601.04064}{{\ttfamily
  arXiv:1601.04064 [gr-qc]}}.

\bibitem{Pan:2025psn}
J.~Pan and G.~Ye, ``{Non-minimally coupled gravity constraints from DESI DR2
  data},'' \href{http://arxiv.org/abs/2503.19898}{{\ttfamily arXiv:2503.19898
  [astro-ph.CO]}}.

\bibitem{Torrado:2020dgo}
J.~Torrado and A.~Lewis, ``{Cobaya: Code for Bayesian Analysis of hierarchical
  physical models},''
  \href{http://dx.doi.org/10.1088/1475-7516/2021/05/057}{{\em JCAP} {\bfseries
  05} (2021) 057}, \href{http://arxiv.org/abs/2005.05290}{{\ttfamily
  arXiv:2005.05290 [astro-ph.IM]}}.

\bibitem{2019ascl.soft10019T}
J.~{Torrado} and A.~{Lewis}, ``{Cobaya: Bayesian analysis in cosmology}.''
  Astrophysics source code library, record ascl:1910.019, Oct., 2019.

\bibitem{Hu:2013twa}
B.~Hu, M.~Raveri, N.~Frusciante, and A.~Silvestri, ``{Effective Field Theory of
  Cosmic Acceleration: an implementation in CAMB},''
  \href{http://dx.doi.org/10.1103/PhysRevD.89.103530}{{\em Phys. Rev. D}
  {\bfseries 89} no.~10, (2014) 103530},
  \href{http://arxiv.org/abs/1312.5742}{{\ttfamily arXiv:1312.5742
  [astro-ph.CO]}}.

\bibitem{Raveri:2014cka}
M.~Raveri, B.~Hu, N.~Frusciante, and A.~Silvestri, ``{Effective Field Theory of
  Cosmic Acceleration: constraining dark energy with CMB data},''
  \href{http://dx.doi.org/10.1103/PhysRevD.90.043513}{{\em Phys. Rev. D}
  {\bfseries 90} no.~4, (2014) 043513},
  \href{http://arxiv.org/abs/1405.1022}{{\ttfamily arXiv:1405.1022
  [astro-ph.CO]}}.

\bibitem{Lewis:1999bs}
A.~Lewis, A.~Challinor, and A.~Lasenby, ``{Efficient computation of CMB
  anisotropies in closed FRW models},''
  \href{http://dx.doi.org/10.1086/309179}{{\em Astrophys. J.} {\bfseries 538}
  (2000) 473--476}, \href{http://arxiv.org/abs/astro-ph/9911177}{{\ttfamily
  arXiv:astro-ph/9911177}}.

\bibitem{Howlett:2012mh}
C.~Howlett, A.~Lewis, A.~Hall, and A.~Challinor, ``{CMB power spectrum
  parameter degeneracies in the era of precision cosmology},''
  \href{http://dx.doi.org/10.1088/1475-7516/2012/04/027}{{\em JCAP} {\bfseries
  04} (2012) 027}, \href{http://arxiv.org/abs/1201.3654}{{\ttfamily
  arXiv:1201.3654 [astro-ph.CO]}}.

\bibitem{Efstathiou:2019mdh}
G.~Efstathiou and S.~Gratton, ``{A Detailed Description of the CamSpec
  Likelihood Pipeline and a Reanalysis of the Planck High Frequency Maps},''
  \href{http://arxiv.org/abs/1910.00483}{{\ttfamily arXiv:1910.00483
  [astro-ph.CO]}}.

\bibitem{Carron:2022eyg}
J.~Carron, M.~Mirmelstein, and A.~Lewis, ``{CMB lensing from Planck
  PR4~maps},'' \href{http://dx.doi.org/10.1088/1475-7516/2022/09/039}{{\em
  JCAP} {\bfseries 09} (2022) 039},
  \href{http://arxiv.org/abs/2206.07773}{{\ttfamily arXiv:2206.07773
  [astro-ph.CO]}}.

\bibitem{DES:2024jxu}
{\bfseries DES} Collaboration, T.~M.~C. Abbott {\em et~al.}, ``{The Dark Energy
  Survey: Cosmology Results with \ensuremath{\sim}1500 New High-redshift Type
  Ia Supernovae Using the Full 5 yr Data Set},''
  \href{http://dx.doi.org/10.3847/2041-8213/ad6f9f}{{\em Astrophys. J. Lett.}
  {\bfseries 973} no.~1, (2024) L14},
  \href{http://arxiv.org/abs/2401.02929}{{\ttfamily arXiv:2401.02929
  [astro-ph.CO]}}.

\bibitem{Efstathiou:2024xcq}
G.~Efstathiou, ``{Evolving Dark Energy or Supernovae Systematics?},''
  \href{http://arxiv.org/abs/2408.07175}{{\ttfamily arXiv:2408.07175
  [astro-ph.CO]}}.

\bibitem{Peng:2025nez}
Z.-Y. Peng and Y.-S. Piao, ``{Dark energy and lensing anomaly in Planck CMB
  data},'' \href{http://arxiv.org/abs/2502.04641}{{\ttfamily arXiv:2502.04641
  [astro-ph.CO]}}.

\bibitem{Huang:2025som}
L.~Huang, R.-G. Cai, and S.-J. Wang, ``{The DESI 2024 hint for dynamical dark
  energy is biased by low-redshift supernovae},''
  \href{http://arxiv.org/abs/2502.04212}{{\ttfamily arXiv:2502.04212
  [astro-ph.CO]}}.

\bibitem{Gialamas:2024lyw}
I.~D. Gialamas, G.~H\"utsi, K.~Kannike, A.~Racioppi, M.~Raidal, M.~Vasar, and
  H.~Veerm\"ae, ``{Interpreting DESI 2024 BAO: Late-time dynamical dark energy
  or a local effect?},''
  \href{http://dx.doi.org/10.1103/PhysRevD.111.043540}{{\em Phys. Rev. D}
  {\bfseries 111} no.~4, (2025) 043540},
  \href{http://arxiv.org/abs/2406.07533}{{\ttfamily arXiv:2406.07533
  [astro-ph.CO]}}.

\bibitem{Park:2024pew}
C.-G. Park, J.~de~Cruz~P\'erez, and B.~Ratra, ``{Is the $w_0w_a$CDM
  cosmological parameterization evidence for dark energy dynamics partially
  caused by the excess smoothing of Planck CMB anisotropy data?},''
  \href{http://arxiv.org/abs/2410.13627}{{\ttfamily arXiv:2410.13627
  [astro-ph.CO]}}.

\bibitem{Park:2024vrw}
C.-G. Park, J.~de~Cruz~P\'erez, and B.~Ratra, ``{Using non-DESI data to confirm
  and strengthen the DESI 2024 spatially flat w0waCDM cosmological
  parametrization result},''
  \href{http://dx.doi.org/10.1103/PhysRevD.110.123533}{{\em Phys. Rev. D}
  {\bfseries 110} no.~12, (2024) 123533},
  \href{http://arxiv.org/abs/2405.00502}{{\ttfamily arXiv:2405.00502
  [astro-ph.CO]}}.

\bibitem{Park:2025azv}
C.-G. Park and B.~Ratra, ``{Is excess smoothing of Planck CMB ansiotropy data
  partially responsible for evidence for dark energy dynamics in other
  $w(z)$CDM parametrizations?},''
  \href{http://arxiv.org/abs/2501.03480}{{\ttfamily arXiv:2501.03480
  [astro-ph.CO]}}.

\bibitem{Silva:2025hxw}
E.~Silva, M.~A. Sabogal, M.~S. Souza, R.~C. Nunes, E.~Di~Valentino, and
  S.~Kumar, ``{New Constraints on Interacting Dark Energy from DESI DR2 BAO
  Observations},'' \href{http://arxiv.org/abs/2503.23225}{{\ttfamily
  arXiv:2503.23225 [astro-ph.CO]}}.

\bibitem{Scherer:2025esj}
M.~Scherer, M.~A. Sabogal, R.~C. Nunes, and A.~De~Felice, ``{Challenging
  $\Lambda$CDM: 5$\sigma$ Evidence for a Dynamical Dark Energy Late-Time
  Transition},'' \href{http://arxiv.org/abs/2504.20664}{{\ttfamily
  arXiv:2504.20664 [astro-ph.CO]}}.

\bibitem{Ye:2024zpk}
G.~Ye, ``{Bridge the Cosmological Tensions with Thawing Gravity},''
  \href{http://arxiv.org/abs/2411.11743}{{\ttfamily arXiv:2411.11743
  [astro-ph.CO]}}.

\bibitem{Wolf:2024stt}
W.~J. Wolf, P.~G. Ferreira, and C.~Garc\'\i{}a-Garc\'\i{}a, ``{Matching current
  observational constraints with nonminimally coupled dark energy},''
  \href{http://dx.doi.org/10.1103/PhysRevD.111.L041303}{{\em Phys. Rev. D}
  {\bfseries 111} no.~4, (2025) L041303},
  \href{http://arxiv.org/abs/2409.17019}{{\ttfamily arXiv:2409.17019
  [astro-ph.CO]}}.

\bibitem{Langlois:2015cwa}
D.~Langlois and K.~Noui, ``{Degenerate higher derivative theories beyond
  Horndeski: evading the Ostrogradski instability},''
  \href{http://dx.doi.org/10.1088/1475-7516/2016/02/034}{{\em JCAP} {\bfseries
  02} (2016) 034}, \href{http://arxiv.org/abs/1510.06930}{{\ttfamily
  arXiv:1510.06930 [gr-qc]}}.

\bibitem{BenAchour:2016cay}
J.~Ben~Achour, D.~Langlois, and K.~Noui, ``{Degenerate higher order
  scalar-tensor theories beyond Horndeski and disformal transformations},''
  \href{http://dx.doi.org/10.1103/PhysRevD.93.124005}{{\em Phys. Rev. D}
  {\bfseries 93} no.~12, (2016) 124005},
  \href{http://arxiv.org/abs/1602.08398}{{\ttfamily arXiv:1602.08398 [gr-qc]}}.

\bibitem{Creminelli:2017sry}
P.~Creminelli and F.~Vernizzi, ``{Dark Energy after GW170817 and GRB170817A},''
  \href{http://dx.doi.org/10.1103/PhysRevLett.119.251302}{{\em Phys. Rev.
  Lett.} {\bfseries 119} no.~25, (2017) 251302},
  \href{http://arxiv.org/abs/1710.05877}{{\ttfamily arXiv:1710.05877
  [astro-ph.CO]}}.

\bibitem{Langlois:2017dyl}
D.~Langlois, R.~Saito, D.~Yamauchi, and K.~Noui, ``{Scalar-tensor theories and
  modified gravity in the wake of GW170817},''
  \href{http://dx.doi.org/10.1103/PhysRevD.97.061501}{{\em Phys. Rev. D}
  {\bfseries 97} no.~6, (2018) 061501},
  \href{http://arxiv.org/abs/1711.07403}{{\ttfamily arXiv:1711.07403 [gr-qc]}}.

\bibitem{Lewis:2019xzd}
A.~Lewis, ``{GetDist: a Python package for analysing Monte Carlo samples},''
  \href{http://arxiv.org/abs/1910.13970}{{\ttfamily arXiv:1910.13970
  [astro-ph.IM]}}.

\bibitem{zenodo_data}
\url{https://doi.org/10.5281/zenodo.15297273}.

\bibitem{Scolnic:2021amr}
D.~Scolnic {\em et~al.}, ``{The Pantheon+ Analysis: The Full Data Set and
  Light-curve Release},''
  \href{http://dx.doi.org/10.3847/1538-4357/ac8b7a}{{\em Astrophys. J.}
  {\bfseries 938} no.~2, (2022) 113},
  \href{http://arxiv.org/abs/2112.03863}{{\ttfamily arXiv:2112.03863
  [astro-ph.CO]}}.

\end{thebibliography}\endgroup


\pagebreak
\widetext
\begin{center}
\textbf{\large Supplemental Material for ``NEC violation and `beyond Horndeski' physics in light of DESI DR2"}
\end{center}
\setcounter{equation}{0}
\setcounter{figure}{0}
\setcounter{table}{0}
\setcounter{page}{1}
\makeatletter
\renewcommand{\theequation}{S\arabic{equation}}
\renewcommand{\thefigure}{S\arabic{figure}}

\section{The covariant form of $\delta g^{00}R^{(3)}$}\label{Sec:cova0327}

The quadratic EFT action writes (see e.g., \cite{Gleyzes:2013ooa}):
\ba
S&=&\int
{\rm d}^4 x \sqrt{-g}\Big[ {M_p^2\over2} f(t)R-\Lambda(t)-c(t)g^{00}
\nn\\
&\,&\qquad\qquad\quad +{M_2^4(t)\over2}(\delta g^{00})^2-{m_3^3(t)\over2}\delta
K\delta g^{00}
-m_4^2(t)\lf( \delta K^2-\delta K_{\mu\nu}\delta
K^{\mu\nu} \rt)
\nn\\
&\,&\qquad\qquad\quad  + {\tilde{m}_4^2(t)\over
	2}\delta g^{00}R^{(3)}\Big] +S_{\rm m}[g_{\mu\nu},\psi_{\rm m}]\,,
\label{action01}
\ea
where higher-order spatial derivative terms are neglected. We have defined $\delta g^{00}=g^{00}+1$, $\delta K_{\mu\nu}=K_{\mu\nu}-h_{\mu\nu}H$, where $H$ is the Hubble parameter, $h_{\mu \nu} = g_{\mu \nu}+n_{\mu} n_{\nu}$ is the induced metric, $n^\mu$ is the unit normal vector of the constant
time hypersurfaces, $K_{\mu\nu}$ is the extrinsic curvature. $R^{(3)}$ is the induced three-dimensional Ricci scalar, and $S_{\rm m}$ represents the action of matter sector, which is minimally coupled to the metric $g_{\mu\nu}$. The EFT coefficients $\{f, \Lambda,c,M_2^4,m_3^3,m_4^2,\tilde{m}_4^2\}$ before the operators are free functions of time and the power only indicates mass dimension but not the sign of the coefficients, e.g. $\tilde{m}_4^2<0$ is allowed. The EFT coefficients $f$, $\Lambda$, $c$, $M_2^4$, $m_3^3$, $m_4^2$, and $\tilde{m}_4^2$ can generally be time-dependent, allowing them to broadly encompass various commonly studied modified gravity theories.

In unitary gauge, $\phi=\phi(t)$, we have \be \delta
g^{00}={X\over \dot{\phi}^2(t)}+1={X\over f_2(t(\phi))}+1,
\label{g00}\ee  where $X=\phi_{\mu} \phi^{\mu}$,
$\phi_\mu=\nabla_\mu\phi$ and $\phi^\mu=\nabla^\mu\phi$. The three-dimensional Ricci scalar
$R^{(3)}$ can be obtained as
\ba \label{covaR3}
R^{(3)}&=& R-{\phi_{\mu\nu}\phi^{\mu\nu}-(\Box \phi)^2\over X}
+{2\phi^\mu\phi_{\mu\nu}\phi^{\nu\sigma}\phi_\sigma \over
X^2}-{2\phi^\mu \phi_{\mu\nu}\phi^\nu \Box \phi\over X^2}
+{2(\phi^\nu_{~\nu\mu}\phi^\mu -\phi_{\nu~\mu}^{~\mu~}
\phi^\nu)\over X}\,,
\ea
where $\phi_{\mu\nu}=\nabla_\nu\nabla_\mu\phi$ and
$\phi^\nu_{~\nu\mu}=\nabla_\mu\nabla_\nu \nabla^\nu\phi$. It can be verified that the right-hand side of Eq. (\ref{covaR3}) vanishes at the background level.

Following \cite{Cai:2017dyi}, we define $S_{\delta g^{00} R^{(3)}}=\int d^4x\sqrt{-g} L_{\delta
g^{00} R^{(3)}}$, where
\ba
\label{covaaction}
L_{\delta g^{00} R^{(3)}}& =& {f_1(\phi)\over 2}\delta g^{00} R^{(3)}\nn\\
&=& {F\over 2}R - {X\over2}\int F_{\phi\phi}d \ln X
-\lf(F_\phi+\int {F_\phi \over 2}d\ln X\rt)\Box\phi \nn\\
&\,& +{F\over 2X}\lf[\phi_{\mu\nu}\phi^{\mu\nu}-(\Box
\phi)^2\rt]
-{F-2X F_X\over
X^2}\lf[\phi^\mu\phi_{\mu\rho}\phi^{\rho\nu}\phi_\nu-(\Box
\phi)\phi^\mu\phi_{\mu\nu}\phi^\nu\rt]
\ea
can be obtained after some integration by parts,
$F(\phi,X)=f_1\lf(1+{X\over f_2}\rt)$ has the dimension of mass squared, $f_2(\phi)$ is defined in
(\ref{g00}), and the total derivative terms have been
discarded.

A simple covariant action that can be used to achieve fully stable NEC violation can be written as \cite{Cai:2017dyi}
\be S=\int d^4x\sqrt{-g}\lf[{M_p^2\over2}R+P(\phi,X) + {\cal L}_{\delta g^{00} R^{(3)}} \rt]\,. \label{eq:act-01}
\ee
This action belongs to the ``beyond Horndeski'' theory (more precisely, the GLPV theory \cite{Gleyzes:2014dya,Gleyzes:2014qga}).

The integral form in Eq. (\ref{covaaction}) may seem confusing. Therefore, the action (\ref{eq:act-01}) can also be further generalized to the following form:
\ba &\,&S=\int d^4x\sqrt{-g}\Big\{{M_p^2+F(\phi,X)\over2}R+P(\phi,X)+G(\phi,X)\Box\phi \nn\\
&\,&\qquad +{F\over 2X}\lf[\phi_{\mu\nu}\phi^{\mu\nu}-(\Box
\phi)^2\rt]
- {F-2X F_X\over
X^2}\lf[\phi^\mu\phi_{\mu\rho}\phi^{\rho\nu}\phi_\nu-(\Box
\phi)\phi^\mu\phi_{\mu\nu}\phi^\nu\rt]\Big\}\,. \label{eq:act-02}
\ea
It is not difficult to verify that Eq. (\ref{eq:act-02}) still falls within the regime of the GLPV theory.


\section{Comparison with different SNIa dataset}\label{apdx:pp}

\begin{figure}[htbp]
    \includegraphics[width=0.44\textwidth]{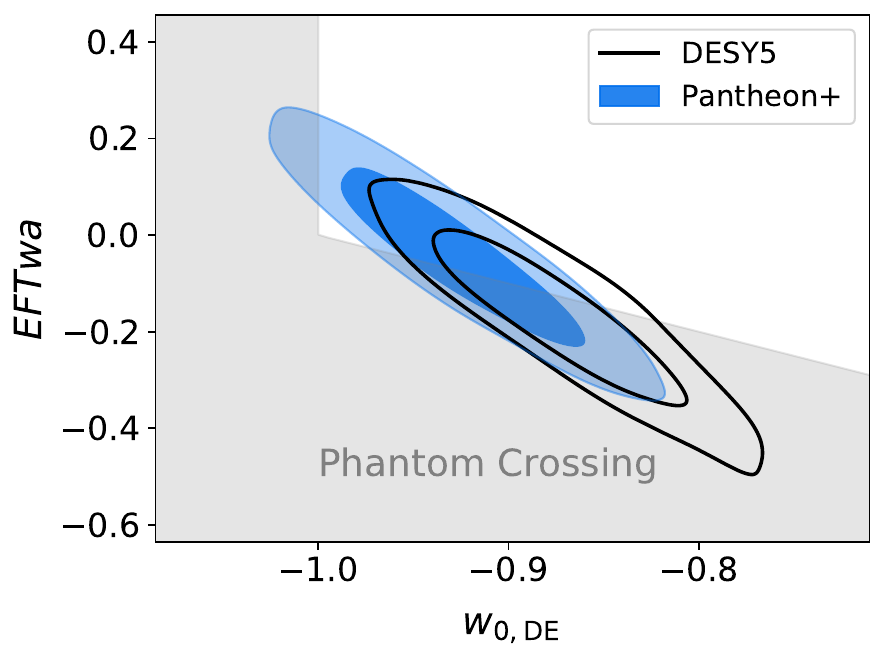}
    \includegraphics[width=0.48\textwidth]{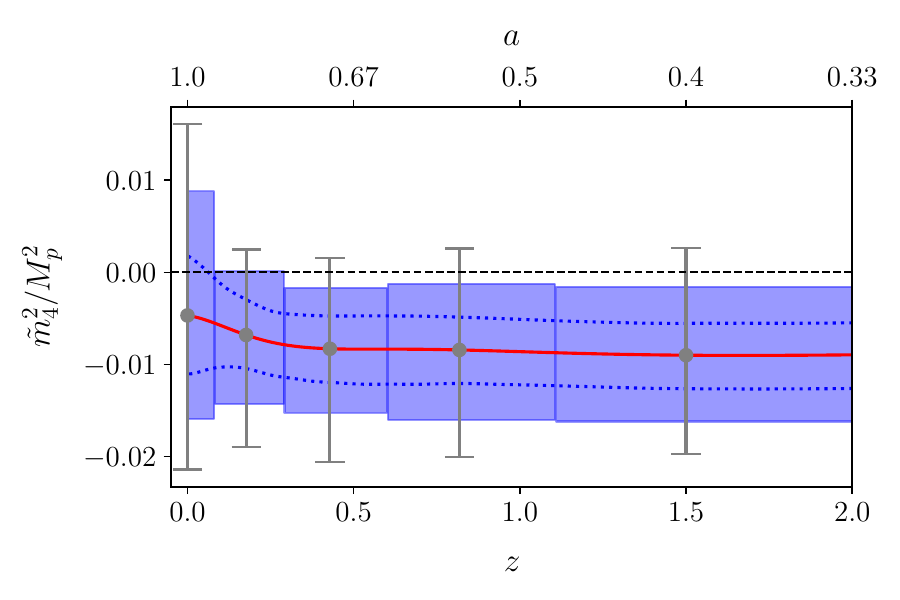}
    \caption{The same model given by Eq. (13) in the main text, with both the $(\delta g^{00})^2$ and $\delta g^{00}R^{(3)}$ operators turned on, and the same plotting style as in Fig. 1, but replacing the SNIa dataset with Pantheon+.}
    \label{fig:pp}
\end{figure}

It is known that using different SNIa data will produce slightly different posterior results about dynamical DE. Here, we perform the same analysis as in the main text but only replace the DESY5 SNIa data with Pantheon+ \cite{Scolnic:2021amr}. Fig. \ref{fig:pp} compares the posterior result, which is statistically consistent.


 \end{document}